\documentclass[twocolumn, 12pt]{aastex631}

\usepackage{placeins}
\usepackage{afterpage}
\usepackage{lipsum}
\usepackage{booktabs}
\usepackage{longtable}
\usepackage{amsmath}
\usepackage{flafter}
\usepackage{xcolor}
\usepackage{url}

\urldef\Vegaurl\url{https://pysynphot.readthedocs.io/en/latest/spectrum.html#pysynphot-vega-spec}

\newcommand{\DeltaPaBeta}{9.8~mag}
\newcommand{\DeltaPaBetaExtended}{9.6~mag}
\newcommand{\JcontPaBetaColor}{0.16~$\pm$~0.26~mag}
\newcommand{\Jcontsynthval}{5.96~$\pm$~0.37~mag}
\newcommand{\PaBetaA}{5.80~$\pm$~0.44~mag}
\newcommand{\PaBetab}{15.60~$\pm$~0.44~mag}
\newcommand{\PaBetabExtended}{15.40~$\pm$~0.44~mag}
\newcommand{\FluxDensityMean}{1.44~$\pm$~0.59$\times$10$^{-15}$~W~m$^{-2}~\mu$m$^{-1}$}
\newcommand{\FluxDensityMeanExtended}{1.73~$\pm$~0.71$\times$10$^{-15}$~W~m$^{-2}~\mu$m$^{-1}$}
\newcommand{\FluxDensityLimit}{2.61$\times$10$^{-15}$~W~m$^{-2}~\mu$m$^{-1}$}
\newcommand{\FluxDensityLimitExtended}{3.14$\times$10$^{-15}$~W~m$^{-2}~\mu$m$^{-1}$}
\newcommand{\EWstar}{--94~$\pm$~64~\AA}

\newcommand{\EWcompanion}{--11~\AA}
\newcommand{\EWcompanionExtended}{--54~\AA}

\newcommand{\PA}{184.2$\degr$}
\newcommand{\sep}{0.60$''$}
\newcommand{\SNR}{S/N}
\newcommand{\PaBmag}{Pa$\beta$}
\newcommand{\PaBfilter}{\emph{Pa}$\beta$}

\newcommand{\Jcontfilter}{\emph{J}$_{cont}$}
\newcommand{\Jcontsyn}{$J_{cont,syn}$}
\newcommand{\JcontPaBcolorLabel}{\emph{J}$_{cont}$--\emph{Pa}$\beta$}

\shorttitle{Deep \PaBmag\ Imaging of AB~Aur~b}
\shortauthors{Biddle et al.}

\begin{document}

\title{Deep Pa$\beta$ Imaging of the Candidate Accreting Protoplanet AB Aur b}

\author[0000-0003-2646-3727]{Lauren I. Biddle}
\affiliation{Department of Astronomy, University of Texas at Austin, 2515 Speedway, Stop C1400, Austin, TX 78712, USA}
\email{lbiddle@utexas.edu}

\author[0000-0003-2649-2288]{Brendan P. Bowler}
\affiliation{Department of Astronomy, University of Texas at Austin, 2515 Speedway, Stop C1400, Austin, TX 78712, USA}

\author[0000-0003-2969-6040]{Yifan Zhou}
\affiliation{Department of Astronomy, 530 McCormick Rd, Charlottesville, VA 22904, USA}

\author[0000-0003-4557-414X]{Kyle Franson}
\altaffiliation{NSF Graduate Research Fellow}
\affiliation{Department of Astronomy, University of Texas at Austin, 2515 Speedway, Stop C1400, Austin, TX 78712, USA}

\author[0000-0002-3726-4881]{Zhoujian Zhang}
\affiliation{Department of Astronomy \& Astrophysics, University of California, Santa Cruz, 1156 High St., Santa Cruz, CA 95064, USA}
\affiliation{Department of Astronomy, University of Texas at Austin, 2515 Speedway, Stop C1400, Austin, TX 78712, USA}




\begin{abstract}
Giant planets grow by accreting gas through circumplanetary disks, but little is known about the timescale and mechanisms involved in the planet assembly process because few accreting protoplanets have been discovered. Recent visible and infrared (IR) imaging revealed a potential accreting protoplanet within the transition disk around the young intermediate-mass Herbig Ae star, AB~Aurigae (AB~Aur). Additional imaging in H$\alpha$ probed for accretion and found agreement between the line-to-continuum flux ratio of the star and companion, raising the possibility that the emission source could be a compact disk feature seen in scattered starlight. We present new deep Keck/NIRC2 high-contrast imaging of AB~Aur to characterize emission in \PaBmag, another accretion tracer less subject to extinction. Our narrow band observations reach a 5$\sigma$ contrast of 9.6 mag at 0$\farcs$6, but we do not detect significant emission at the expected location of the companion, nor from other any other source in the system. Our upper limit on \PaBmag\ emission suggests that if AB Aur b is a protoplanet, it is not heavily accreting or accretion is stochastic and was weak during the observations.

\end{abstract}



\section{Introduction} \label{sec:Introduction}

Direct observations of planetary assembly with high-contrast imaging offers a promising pathway to establish the timescale, location, and mechanisms of giant planet formation. Planets accrete mass from a circumplanetary disk, which may be continually fed by a circumstellar disk during an embedded phase. Recent modeling efforts have informed observations of planetary-mass accretion signatures (e.g., \citealt{Thanathibodee2019, Szulagyi2020, Aoyama2020, Marleau2022, Choksi2022}), predicting increased favorability of detection at wavelengths with strong accretion luminosity such as line emission from infalling hydrogen gas and excess emission in the UV continuum from accretion hot spots. Ultraviolet (UV) to near-infrared (NIR) imaging of a handful of wide accreting substellar companions show signatures that are consistent with theoretical predictions (e.g., CT~Cha~B, \citealt{Schmidt2008}; DH~Tau~B, \citealt{Zhou2014, vanHolstein2021}; GQ~Lup~B, \citealt{Zhou2014, Stolker2021, Demars2023}; GSC~06214-00210~B, \citealt{Seifahrt2007, Bowler2011, Zhou2014, vanHolstein2021, Demars2023}; SR~12~C, \citealt{Santamaria-Miranda2018})---most notably strong H$\alpha$ and Pa$\beta$ emission, and relatively bright UV emission.

However, protoplanets still embedded in the circumstellar disk pose a challenge; distinguishing disk substructures such as concentrated unresolved clumps from \textit{bona fide} accreting protoplanets is complex, and many claimed protoplanet candidates identified at thermal wavelengths and in H$\alpha$ emission have been debated (e.g., LkCa~15~bcd \citealt{Thalmann2016, Mendigutia2018, Currie2019}; HD~100546~bc \citealt{Rameau2017, Follette2017, Sissa2018}; HD~169142~bc \citealt{Biller2014, Pohl2017, Ligi2018}). PDS 70 b and c \citep{Keppler2018, Wagner2018, Haffert2019} are the only uncontested accreting protoplanets. Recently, visible and IR imaging revealed a source of concentrated emission in the transition disk surrounding the young ($\approx$4--6 Myr; \citealt{GuzmanDiaz2021, Zhou2022}) accreting A0 star AB~Aurigae \citep{Currie2022}, which was interpreted as an embedded protoplanet at a projected separation of 0.6$''$ ($\approx100$~au). This is near the location where a giant planet was predicted by \citet{Tang2017} via modeling the disk’s inner substructure. The source is spatially resolved in multiple datasets and may be tracing circumplanetary disk structure, an envelope, circumstellar disk substructure, or a combination of these.

Protoplanets that are embedded in transition disks are predicted to experience persistent accretion (e.g., \citealt{Tanigawa2012, Szulagyi2014}) and will therefore exhibit strong H$\alpha$ emission from gas that is excited or ionized by intense radiation from the accretion post-shock region \citep{Aoyama2018}. If the surrounding envelope is not optically thick, then some or all of this emission will escape. However, the central star in transition disk systems can itself accrete from an inner disk and will also exhibit hydrogen emission. In these cases, the spectrum of a bright unresolved dust feature in the disk seen in scattered light could resemble an accreting planet.

In the case of AB Aur, the host is accreting and shows strong and variable line emission (e.g., \citealt{HarringtonKuhn2007, Costigan2014}). For this reason, confirming the protoplanetary nature of AB~Aur~b requires careful disambiguation of its own emission from that of surrounding disk features seen in scattered light. Any difference in line strength relative to the continuum between AB~Aur~A and b would support the hypothesis that AB~Aur~b is a protoplanet and would point to emission being locally generated. On the contrary, a consistent line strength would suggest that AB~Aur~b is actually a compact disk feature seen in scattered starlight.

\citet{Zhou2022} used the Hubble Space Telescope (HST) to search for H$\alpha$ emission from AB~Aur~b using narrow band photometry and found that moderate H$\alpha$ excess is indeed present. In addition, the line-to-continuum ratio of the candidate planet and star are consistent, which calls into question whether the emission originates from planetary accretion or scattered emission. However, the observations were not obtained simultaneously, so the similarity could also be coincidental. New UV and optical HST imaging by \citet{Zhou2023} indicates that at short wavelengths, the emission from AB~Aur~b is dominated by scattered light from the host star.

If AB~Aur~b is an accreting planet, then higher-order emission lines such as \PaBmag\ should also be present. \PaBmag\ is advantageous in searching for distinct accretion signatures from AB~Aur~b because it is less affected by extinction from the disk in comparison to H$\alpha$. In addition, a possible complimentary outcome from a multi-wavelength analysis is a constraint on the line-of-sight dust extinction to the source. This could further inform strategies for future imaging of AB~Aur~b, for instance with the James Webb Space Telescope.  

Here, we report the first constraints on \PaBmag\ emission from AB Aur b from Angular Differential Imaging (ADI) observations obtained with Keck II/NIRC2. We also characterize the stellar NIR emission line properties of AB Aur with moderate-resolution 0.74--2.50~$\mu$m spectroscopy acquired with the NASA Infrared Telescope Facility (IRTF)/SpeX. We outline our observations and data reduction in Section \ref{sec:Observations}. In Section \ref{sec:Results}, we present the results. See Section \ref{sec:Discussion} for our discussion of the resulting contrast curve and constraint on \PaBmag\ line strength. A summary of this work can be found in Section~\ref{sec:summary}.

\section{Observations} \label{sec:Observations}

\subsection{Keck II/NIRC2}\label{sec:Keck}

Adaptive optics images were obtained on UT 2022 September 17 with the NIRC2 camera on the Keck II 10-meter Telescope \citep{Wizinowich2000}. The NIRC2 plate scale is 9.971~$\pm$~0.004 mas px$^{-1}$ and the Position Angle (PA) of each column relative to North is 0.262~$\pm$~0.020$\degr$ \citep{Service2016}. Images were acquired in vertical angle mode. We used the central star AB~Aur~A ($G$ = 7.11 mag; \citealt{GaiaCollaboration2020}) as a natural guide star. The seeing was stable at $\approx0.5''$ throughout the observations. All images were read as a centered 512~$\times$~512 subarray.

We acquired a sequence of 50 images in the \PaBfilter\ narrow band filter\footnote{The NIRC2 narrow band \PaBfilter\ filter transmission profile was digitally extracted from the documentation provided on the NIRC2 instrument page (\url{https://www2.keck.hawaii.edu/inst/nirc2/filters.html}). Filter scans were measured at $\approx$300~K, however the NIRC2 operating temperature is $\approx$50~K, resulting in a wavelength offset of the transmission scan. We apply a wavelength correction $\Delta\lambda$ dictated by the thermal expansion properties of the filter: $\Delta\lambda$ = $\alpha\Delta T$ where $\alpha$ = 2.55$\times$10$^{-5}$~$\mu$m~K$^{-1}$ (R. Campbell, private communication, 2023). For $\Delta T$ = 250~K, $\Delta\lambda$ = --0.0064~$\mu$m.\label{foot:PaBetaFilter}} ($\lambda_{0} = 1.2903$ $\mu$m, $\Delta\lambda = 0.0193$ $\mu$m) using Angular Differential Imaging (ADI; \citealt{Liu2004, Marois2006}). The ADI method requires that the telescope pupil remain fixed on the camera while the field of view (FOV) rotates around the star over time. Each image in the ADI sequence was taken with a 1 sec integration time and 20 coadds for a total time of 20~s per frame. The total angular rotation of the FOV was 57.9$\degr$, and the total on-source exposure time in \PaBfilter\ was 16.7~min.

Because AB~Aur saturates in the ADI frames, we also captured 25 frames in the \PaBfilter\ filter with exposure times of 0.2 sec, and 100 coadds. AB~Aur~A was not saturated in these frames, which enabled us to calibrate the contrast of the deep \PaBfilter\ sequence. In addition, 15 images in the \Jcontfilter\ narrow band filter\footnote{The same correction for the temperature difference between the filter profile scan and the operating temperature was applied to the \Jcontfilter\ filter profile as was carried out for \PaBfilter\ (see Footnote \ref{foot:PaBetaFilter}), except here $\alpha$~=~3.18$\times$10$^{-5}$~$\mu$m~K$^{-1}$ (R. Campbell, private communication, 2023), resulting in a wavelength offset of $\Delta\lambda$ = --0.0080~$\mu$m.} ($\lambda_{0} = 1.2132$~$\mu$m, $\Delta\lambda$~=~0.0198~$\mu$m) were obtained with exposure times of 0.3 sec. The \Jcontfilter\ images allow us to calibrate our measured \PaBfilter\ emission amplitude of the host star relative to the \emph{J}-band continuum.

Additional NIRC2 images were taken on UT 2023 May 3 of the A0V standard star HD~109691 to accurately calibrate the AB~Aur \PaBmag\ line strength for the system throughput response (see Section \ref{subsec:Monochromatic_Flux_Density_Steps}). We took two sequences of 10 exposures each in the \Jcontfilter\ and \PaBfilter\ filters with an integration time of 2 sec with 1 coadd. A 256~$\times$~256 subarray was used for both sequences. The average seeing during the observations was $\approx$0$\farcs$5.

The raw images were cleaned of bad pixels and cosmic rays, then bias-subtracted and flat-fielded. Optical distortions in the images were corrected with a geometric distortion model for the NIRC2 the narrow field mode characterized by \citet{Service2016}. The distortion model was applied to the images using \texttt{Rain}\footnote{\url{github.com/jsnguyen/rain} developed by Jayke S. Nguyen.}, a publicly available Python adaptation of the linear reconstruction software \texttt{Drizzle} \citep{FruchterHook2002}. We aligned and median-combined the unsaturated frames of AB Aur in \PaBfilter\ and \Jcontfilter.

\subsection{IRTF/SpeX}\label{sec:IRTF}

A moderate-resolution ($R \approx 2000$) near-infrared spectrum of AB~Aur~A was acquired from the NASA Infrared Telescope Facility (IRTF) on UT 2022 September 25. We used the SpeX spectrograph \citep{Rayner2003} in the short-wavelength cross-dispersed (SXD) mode with the $0.3~\times~15''$ slit and took 12 exposures with 10~sec each in a standard ABBA pattern. The A0V standard star, HD~31069, was observed for telluric correction within an airmass of 0.06 of AB~Aur~A. We reduced the data using version 4.1 of the Spextool software package\footnote{\url{http://irtfweb.ifa.hawaii.edu/~spex/observer/}} \citep{Cushing2004} and obtained a 0.15~\AA\ dispersion in the wavelength calibration. Our resulting SXD spectrum spans 0.7--2.5~$\mu$m, with a signal-to-noise (S/N) ratio of $\approx440$ per pixel near 1.0~$\mu$m.

\begin{figure*}[t!]
    \centering
    \includegraphics[width=\textwidth]{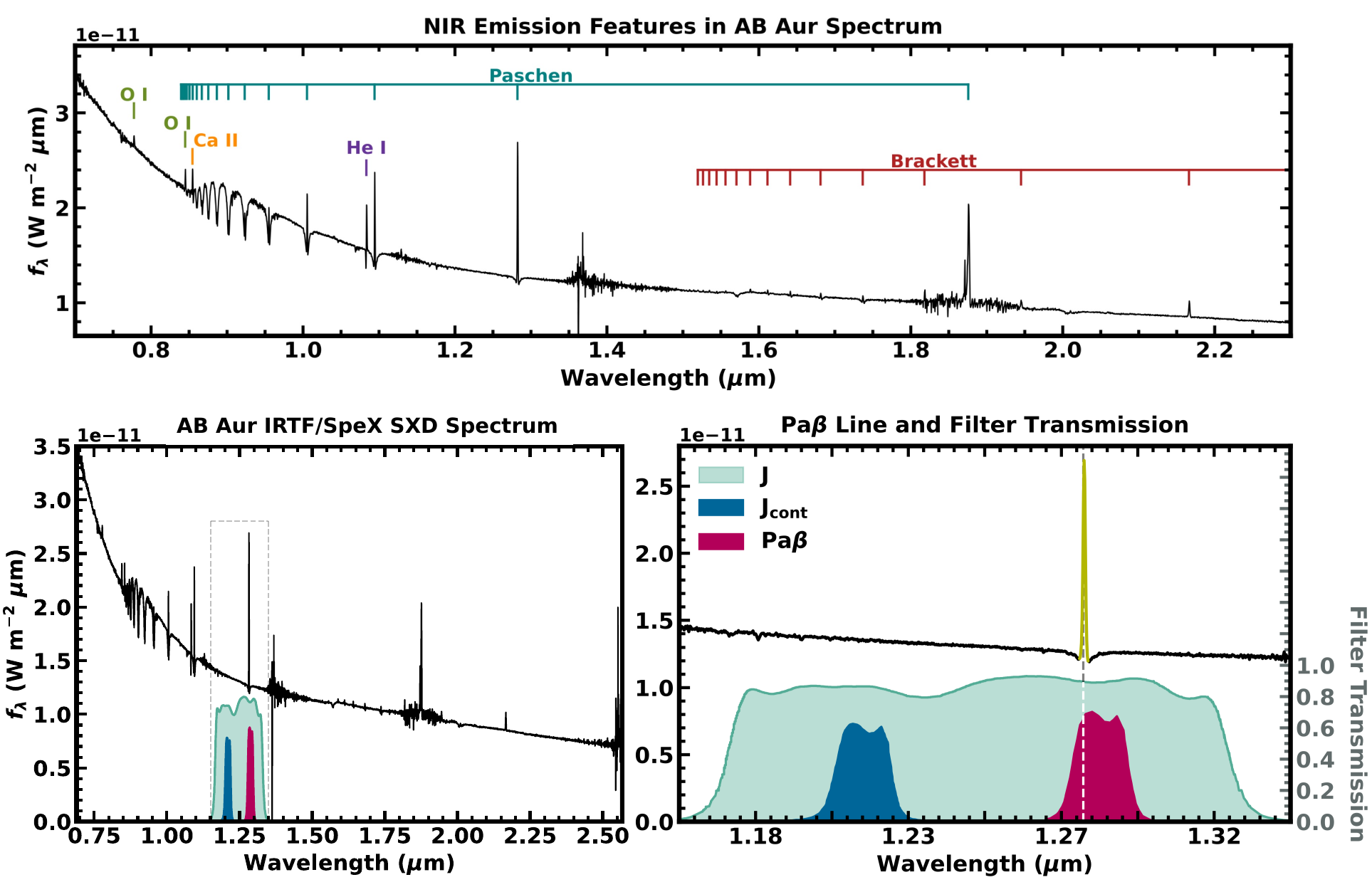}
    \caption{\emph{Upper Panel}: A mapping of the location of emission lines in the NIR IRTF spectrum of AB~Aur~A. \emph{Lower Left}: The entire flux-calibrated IRTF/SpeX SXD spectrum of AB~Aur~A (Section \ref{sec:IRTF}). The spectral ranges $\approx$1.3--1.5~$\mu$m and $\approx$1.7--1.9~$\mu$m fall in telluric regions and have reduced signal to noise. The grey dotted box indicates the spectral region containing the \PaBmag\ line and the NIRC2 filter bandpasses utilized in this work. \emph{Lower Right}: A zoomed-in perspective of the region within the dotted box in the left panel. The light green emission component of the \PaBmag\ line indicates the region that was masked during our calculation of the host star's synthetic \Jcontfilter\ magnitude (\Jcontsyn). The transmission profiles of the relevant NIRC2 filters are displayed below (\emph{J} filter is shown in teal, the \Jcontfilter\ filter is shown in blue, and the \PaBfilter\ filter is shown in red). The vertical dashed line traces \PaBmag\ emission line location.}
    \label{fig:Spectrum}
\end{figure*}

We catalogued the presence and strength of the emission lines in the NIR IRTF SXD spectrum of AB~Aur (Figure \ref{fig:Spectrum}). Many of the hydrogen emission lines occur deep within the line core of a broader absorption feature from the stellar atmosphere, so we compute the Equivalent Width (EW) only in the wavelength region where the line is in emission, adopting a straight line interpolation of the flux at both endpoints of the emission component as the continuum. In addition to emission from the H I Paschen and Brackett series, we also observe O I, Ca II, and the He I $\lambda$1.0830~$\mu$m line. Table \ref{tab:emission_lines} provides identifying information for the prominent emission lines from AB~Aur, the observed vacuum wavelength~$\lambda_{obs}$, as well as their corresponding EW. 

\begin{deluxetable}{lccc}[h!]
\tabletypesize{\footnotesize}
\tablecolumns{2}
\tablewidth{0pt}
\tableheadfrac{0.1}
\tablecaption{Emission Lines in the IRTF/SpeX NIR Spectrum of AB Aur.}
\label{tab:emission_lines}
\tablehead{
    \colhead{Line ID} & \colhead{$\lambda_{obs}$} & \colhead{EW} & \colhead{$\sigma_{\mathrm{EW}}$} \\
    \colhead{} & \colhead{($\mu$m)} & \colhead{(\AA)} & \colhead{(\AA)}
    }
\startdata
O I     &        0.7776 &   -0.33 &        0.06 \\
O I + H I (18--3) &        0.8451 &   -0.62 &        0.05 \\
Ca II + H I (15--3) &        0.8546 &   -0.97 &        0.05 \\
H I  (14--3) &        0.8603 &   -0.09 &        0.03 \\
Ca II + H I  (13--3) &        0.8668 &   -0.32 &        0.04 \\
H I  (12--3) &        0.8754 &   -0.18 &        0.03 \\
H I  (11--3) &        0.8867 &   -0.36 &        0.03 \\
H I  (10--3) &        0.9019 &   -0.42 &        0.03 \\
H I  (9--3) &        0.9233 &   -0.89 &        0.04 \\
H I  (8--3) Pa$\epsilon$ &         0.955 &   -1.51 &        0.06 \\
H I  (7--3) Pa$\delta$ &        1.0054 &   -2.96 &        0.05 \\
He I $\lambda$1.0830 $\mu$m &      1.0837$^{\mathrm{a}}$ &    -2.40 &        0.06 \\
H I  (6--3) Pa$\gamma$ &        1.0944 &   -5.58 &        0.08 \\
H I  (5--3) Pa$\beta$ &        1.2824 &  -11.58 &        0.09 \\
H I  (19--4) &        1.5268 &   -0.17 &        0.05 \\
H I  (18--4) &        1.5349 &   -0.22 &        0.05 \\
H I  (17--4) &        1.5444 &   -0.24 &        0.05 \\
H I  (16--4) &        1.5565 &   -0.19 &        0.05 \\
H I  (15--4) &        1.5708 &   -0.16 &        0.04 \\
H I  (14--4) &        1.5888 &   -0.42 &        0.05 \\
H I  (13--4) &        1.6116 &   -0.51 &        0.06 \\
H I  (12--4) &        1.6414 &    -0.50 &        0.06 \\
H I  (11--4) &        1.6814 &   -0.62 &        0.07 \\
H I  (10--4) &        1.7369 &   -0.82 &        0.09 \\
H I  (4--3) &        1.8759 &  -21.41 &        3.71 \\
H I  (7--4) Br$\gamma$ &        2.1663 &   -3.28 &        0.13 \\
\enddata
\tablecomments{$^{\mathrm{a}}$The He I $\lambda$1.0830$\mu$m line exhibits a clear P-Cygni shape, and the reported $\lambda_{obs}$ signifies the location of the peak of the emission component only.}
\end{deluxetable}


\begin{figure*}[!ht]
    \centering
    \includegraphics[width=\textwidth]{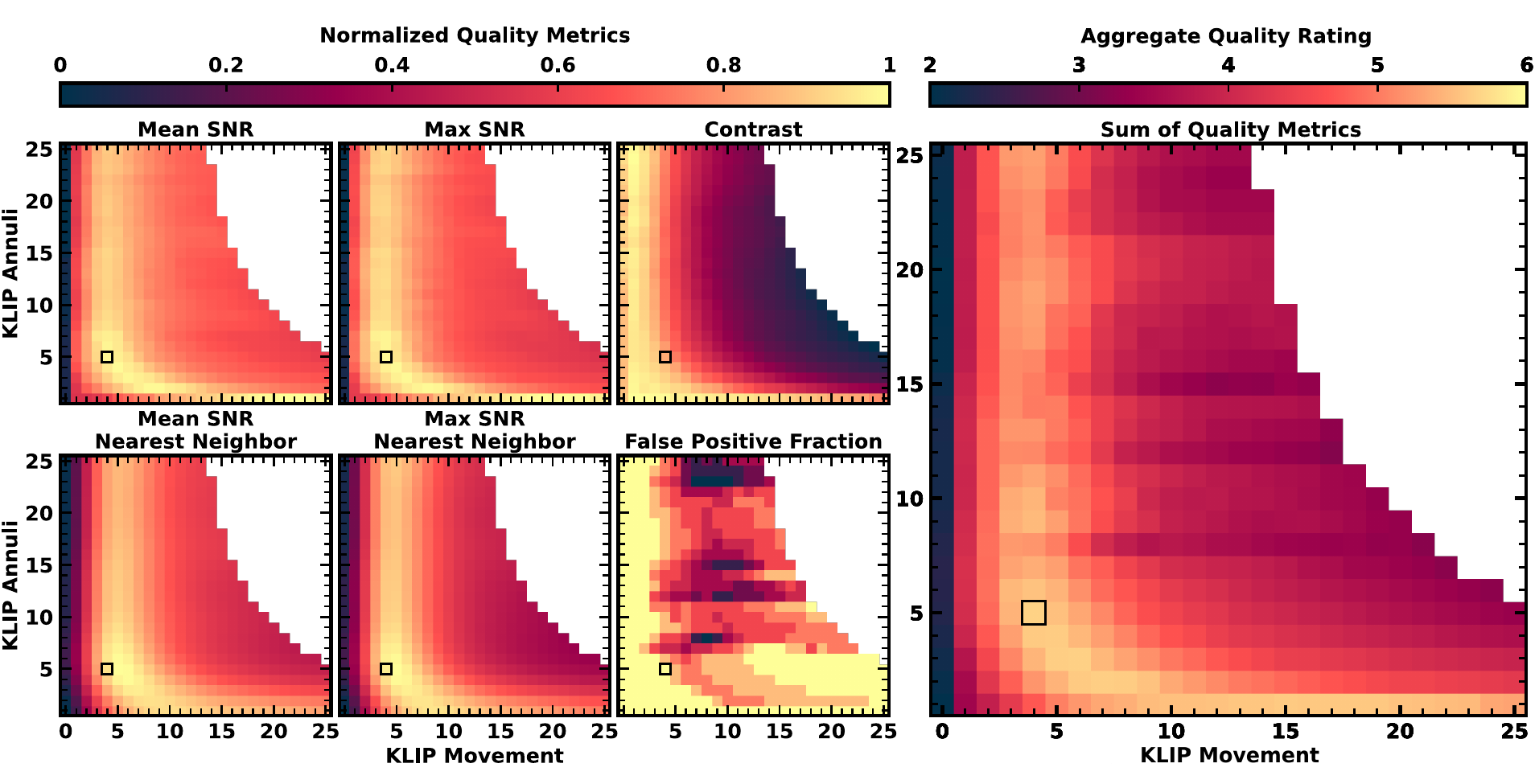}
    \caption{\textit{Left 6 Panels}: Normalized quality maps based on injection-recovery tests for the \texttt{movement} and \texttt{annuli} parameters of the KLIP processing using 20 principle components. From upper-left to lower-right, the metrics presented are: mean \SNR, maximum \SNR, average contrast, mean \SNR\ nearest neighbor, max \SNR\ nearest neighbor, and false-positive fraction. See Section \ref{sec:PSF_Subtraction} for details. Individual maps are normalized and colored according to the respective quality rating (0=worst quality, 1=best quality). These quality metric maps are used to compute the aggregate quality map from which the optimal parameter combinations are chosen. \textit{Right Panel}: The aggregate quality map constructed from the equally weighted summation of all 6 quality metric maps. The cell with the highest rating from the aggregate map is outlined in black in every panel. Cells containing no quality rating have a combination of \texttt{movement} and \texttt{annuli} parameters that leave no reference images for PSF subtraction as a result of the aggressiveness of the rotational mask.}
    \label{fig:Aggregate_Quality_Map}
\end{figure*}

\section{Results}\label{sec:Results}

\subsection{PSF Subtraction}\label{sec:PSF_Subtraction}

We performed PSF subtraction on the 50 long-exposure \PaBfilter\ images from the ADI sequence with \texttt{pyKLIP} software\footnote{\url{https://pyklip.readthedocs.io/en/latest/}} \citep{Wang2015}. The KLIP parameters used to produce the PSF-subtracted image were selected according to a data-driven approach inspired by the ``pyKLIP Parameter Explorer" algorithm (\texttt{pyKLIP-PE}) developed by \citet{Adams2023}, whereby the most optimal KLIP parameters are identified based on several image quality metrics describing the effectiveness of PSF subtraction in an injection-recovery analysis. The primary advantage of this data-driven approach is that it ensures the procedure for PSF subtraction does not rely on a prior significant point source detection and reduces cognitive biases in parameter selection (e.g., \citealt{Follette2022, Adams2023}).


Our approach aims at optimizing the following parameters that were found by \citet{Follette2022} and \citet{Adams2023} to have the greatest impact on PSF subtraction with \texttt{pyKLIP}: \texttt{numbasis} (the number of principal components used to construct the PSF), \texttt{movement} (the minimum number of pixels through which a planet will have rotated between the target and reference images), and \texttt{annuli} (the number of concentric and equal-width annular zones that are analyzed independently by \texttt{pyKLIP}).

The procedure for our adopted algorithm is as follows:
\begin{enumerate}
    
    \item We first inject several synthetic planets into the pre-processed images at PAs increasing sequentially by 85$\degr$ and radial separations of one FWHM (4.5 pix, or 0.044$''$), beginning at a separation of 0.2$''$ outward to 0.75$''$. Each synthetic planet is assigned a flux of 7 times the standard deviation of the local image background; a value slightly greater than the 5$\sigma$ detection threshold, which allows a range of parameter combinations to be tested that will still result in a robust detection.

    \item Next, an individual KLIP reduction is performed for every combination of input parameters \texttt{numbasis} = [5, 10, 15, 20, 25, 30], \texttt{movement} = [1, 2, ..., 24, 25], and \texttt{annuli} = [1, 2, ..., 24, 25], probing a range of aggressive and conservative permutations of PSF subtraction. The parameters held constant across all reductions are the inner working angle (\texttt{IWA}=20 pixels), and the number of subsections used to divide the annular zones (\texttt{subsections}=1).

    \item Then, the point source detection sensitivity for each combination of parameters is represented as a combination of six quality metrics that are determined based on the recovered properties of the synthetic planets. The quality metrics are:
    \begin{itemize}
        \item The average recovered \SNR\ of the injected planets.
        \item The maximum recovered \SNR\ of the injected planets.
        \item The average recovered contrast of the injected planets.
        \item The ``false positive fraction" of the image pixels---the fraction of pixels with a S/N $>$5$\sigma$ within an annulus at the injected planet's separation with thickness $\Delta r$ = 5 pix, or 0.048$''$ (slightly larger than the FWHM of 4.5 pix, to ensure all necessary pixels are counted). The regions containing the candidate and injected planets are first masked before determining this false positive fraction.
        \item The final two ``neighbor quality'' metrics represent the average and maximum \SNR\ metric values listed above. The neighbor quality metrics are obtained by smoothing the average and maximum \SNR\ in \texttt{movement}/\texttt{annuli} space for each KL parameter ``slice''. Neighbor quality metrics are an important consideration because small variations between KLIP parameters should not heavily affect the \SNR\ recovered from a real planet. Smoothing will therefore penalize combinations of \texttt{movement} and \texttt{annuli} parameters that produce drastic variations in \SNR\ among neighboring cells.
    \end{itemize} 

\end{enumerate}

The final selection of \texttt{movement} and \texttt{annuli}, output metrics that were computed for \texttt{numbasis}~=~5, and \texttt{numbasis}~=~30 are averaged, normalized to unity, and summed to produce a single ``aggregate quality map'' (Figure \ref{fig:Aggregate_Quality_Map}). The cell containing the highest aggregate quality metric (a maximum possible value of 6) yields the combination of \texttt{movement} and \texttt{annuli} parameters that is optimal, assuming equal weights for each metric in this framework. For this \PaBfilter\ dataset, the best values are \texttt{movement}=4 and \texttt{annuli}=5.

\begin{figure*}[!ht]
    \centering
    \includegraphics[width=\textwidth]{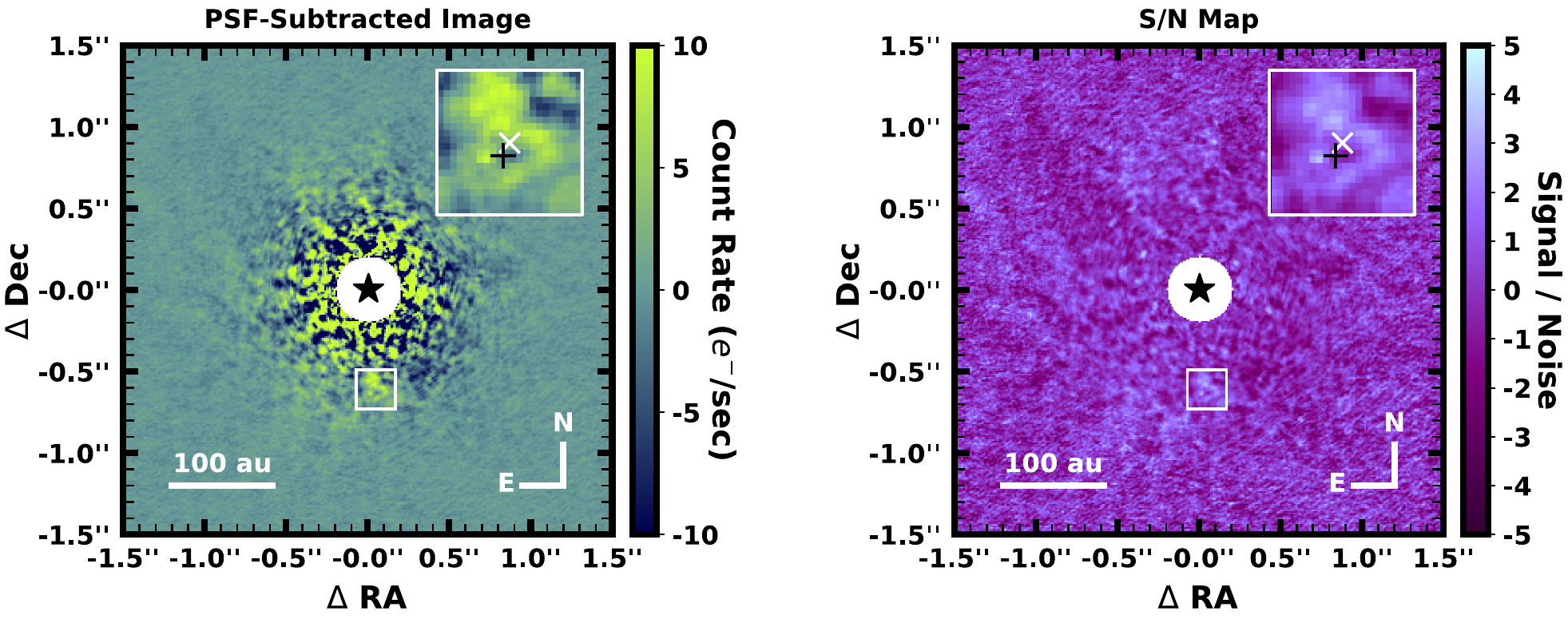}
    \caption{\textit{Left}: The PSF-subtracted image of AB~Aur produced with optimized \texttt{pyKLIP} processing parameters (Section \ref{sec:PSF_Subtraction}). \textit{Right}: The \SNR\ map of the PSF-subtracted image where the \SNR\ of each pixel is computed as the flux divided by the standard deviation of the background in a narrow annulus at that pixel's separation. The inset frame in the upper right portion of both panels provides an enlarged perspective of the region outlined by the white box at the expected location of the companion. Within the inset views, the white ``$\times$'' indicates the most recent location reported by \citet{Currie2022} (sep = \sep, PA = \PA), and the black ``+'' gives the location reported by \citet{Zhou2022} (sep = 0.60$''$, PA = 182.5$\degr$). The \SNR\ within an aperture radius of 3$\times$FWHM at the position of the white ``$\times$'' is 1.4, and the \SNR\ at the position of the black ``+'' is 0.8, both of which are below our threshold of 5$\sigma$ that we use to establish a reliable detection.}
    \label{fig:2panel}
\end{figure*}

To determine the final \texttt{numbasis} component, we repeat this process without averaging any \texttt{numbasis} slices, now producing six aggregate quality maps---one for each principle component that we tested. The optimal \texttt{numbasis} is that which corresponds to the aggregate quality map with the highest metric corresponding to \texttt{movement}=4 and \texttt{annuli}=5, which we determine to be \texttt{numbasis}=20. Ultimately, the final set of the KLIP parameters that produce the highest quality PSF-subtracted image (Figure \ref{fig:2panel}) is \texttt{movement}=4, \texttt{annuli}=5, and \texttt{numbasis}=20.

\subsection{Characterizing Detection Sensitivity in Pa$\beta$}\label{subsec:Sensitivity} 

We conducted injection-recovery tests to quantify the throughput of the KLIP reduction results and calibrate the 5$\sigma$ \PaBfilter\ contrast. Because AB~Aur~b has been spatially resolved in previous detections (e.g., \citealt{Currie2022, Zhou2023}), we perform the injection-recovery test for spatially resolved sources. We also perform the nominal test for unresolved sources to quantify throughput of all potential companions in the field. For both cases, synthetic planets are injected over a series of PAs spaced by 40$\degr$, avoiding overlap with the expected PA of AB~Aur~b ($\approx$180$\degr$). For the spatially resolved case, we construct the template of the injected profile by convolving a 2-dimensional Gaussian kernel with a FWHM equal to the NIRC2 angular resolution at 1.29~$\mu$m (0.044$''$) with with a flat circle with a radius equal to the intrinsic radius of AB~Aur~b's spatial profile of $\approx$0.045$''$ \citep{Currie2022}. The resulting profile is a spatially extended source with a FWHM~=~0.086$''$.

\begin{figure}[!ht]
    \centering
    \includegraphics[width=\columnwidth]{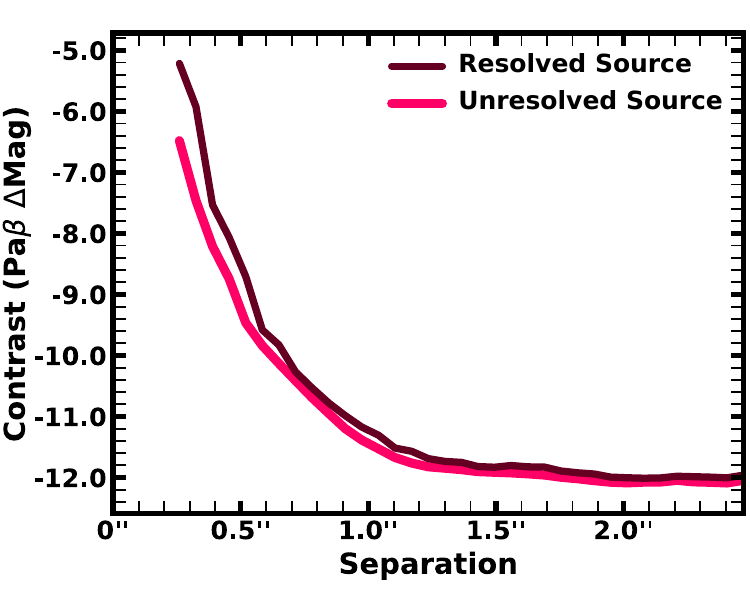}
    \caption{The 5$\sigma$ contrast curve of the PSF-subtracted image of AB~Aur in the NIRC2 narrow band \PaBfilter\ filter computed for an unresolved source (pink) and a resolved source (maroon). Although the expected $\Delta$\emph{J} contrast is $\approx$9.2 \citep{Currie2022}, the relative \PaBmag\ emission line strength between the accreting host star and the candidate companion will impact the contrast of the companion in the \PaBfilter\ filter.}
    \label{fig:contrast_curve}
\end{figure}

We injected the spatially extended source at radial separations equally spaced by 2$\times$FWHM of the injected profile, and each profile was scaled to a peak flux equal to 15 times the standard deviation of the noise within an annulus 1-pixel wide at that planet's separation. These fluxes ensure that the throughput is calculated accurately in post-processing with minimal effects from artifacts that can be created by bright synthetic planets. The throughput was then computed from the average ratio of the injected and recovered fluxes. Then, the \SNR\ map of the PSF-subtracted image was constructed, first masking out a circular region 2$\times$FWHM in diameter centered on the expected location of the companion. For each pixel, we measure the flux in an aperture with a diameter equal to twice the FWHM of the PSF. Next, the aperture-integrated flux was divided by the standard deviation of the fluxes in an annulus covering the remaining azimuthal region, producing a map of the \SNR\ for every pixel in the image (Figure \ref{fig:2panel}). We then derived an initial 5$\sigma$ contrast curve of the post-processed image by calibrating the deep images with unsaturated \PaBfilter\ frames. The location of the candidate was masked during this process, and the $t$-distribution correction was applied to account for small number statistics close to the inner working angle \citep{Mawet2014}. The throughput-corrected 5$\sigma$ contrast curve was computed by dividing the uncorrected contrast value at each resolution element with the throughput at the nearest separation. Figure \ref{fig:contrast_curve} shows the resulting \PaBfilter\ contrast achieved at separations out to $\approx$2.5$''$ with tabulated values of separation vs \PaBfilter\ contrast given in Appendix \ref{appendix:contrast}. 

\subsection{No Detection of Pa$\beta$ Emission}\label{sec:NoCigar}

\begin{figure*}[!ht]
    \centering
    \includegraphics[width=\textwidth]{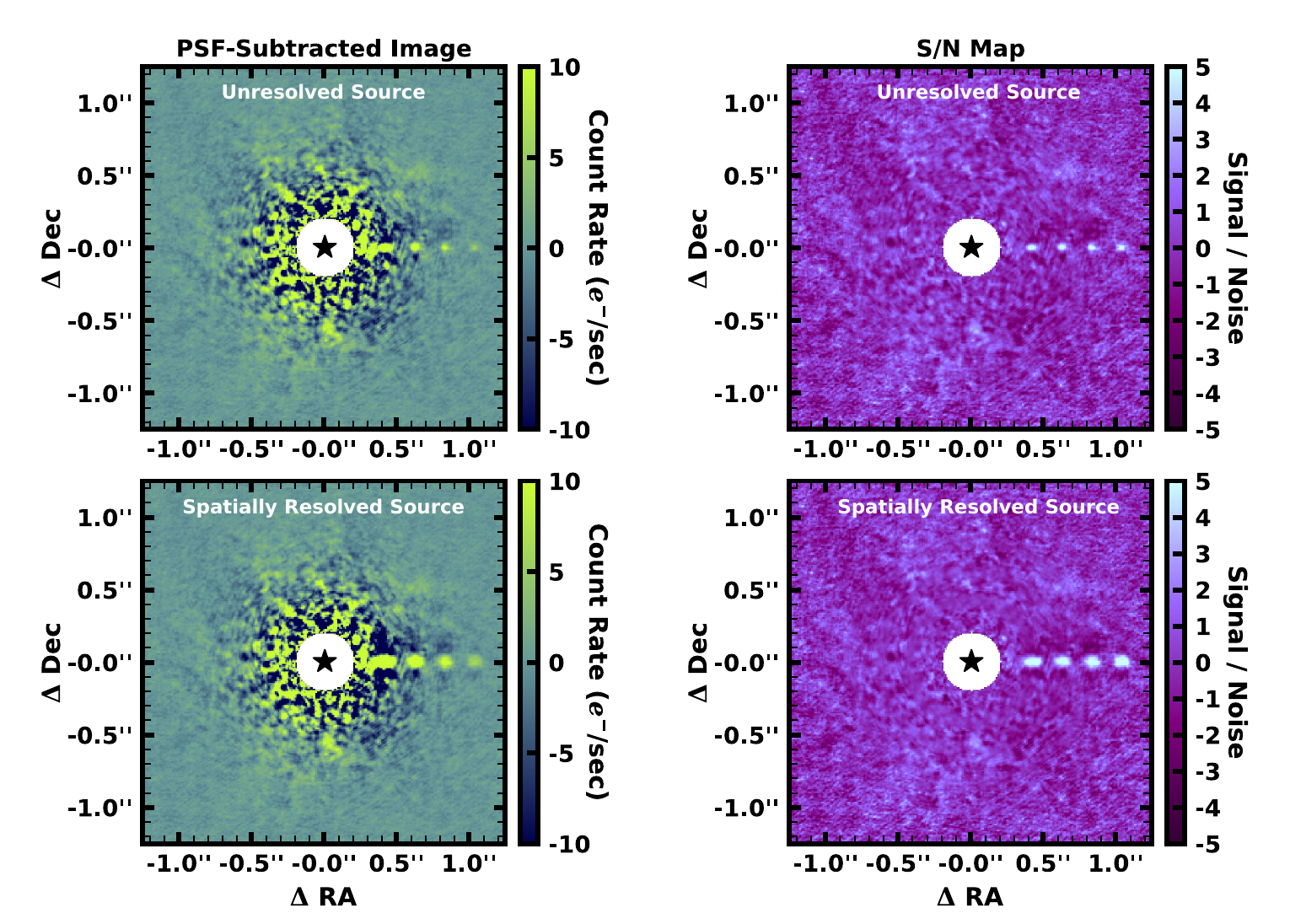}
    \caption{The PSF-Subtracted Image (\textit{left column}) and corresponding S/N Map (\textit{right column}) showing several 5$\sigma$ injected sources that are unresolved (\textit{top row}) and spatially resolved (\textit{bottom row}). The synthetic sources are located at PA = 270$\degr$ with separations 0.4$''$, 0.6$''$, 0.8$''$, and 1.0$''$. The FOV is slightly smaller than that in Figure \ref{fig:2panel} for enhanced visual clarity.}
    \label{fig:2panel_5sigma_injected}
\end{figure*}

The \SNR\ map of the PSF-subtracted image reveals neither a clear spatially resolved source nor an unresolved source with a \SNR\ ratio at the 5$\sigma$ detection threshold anywhere in the 2$\farcs$5 $\times$ 2$\farcs$5 FOV surrounding the star. At the expected location of AB~Aur~b corresponding to the epoch reported by \citet{Currie2022} (UT 2020 October 12: sep = 0.599$''$~$\pm$~0.005$''$, PA = 184.2~$\pm$~0.6$\degr$), the \SNR\ = 1.4. At the location corresponding to the epoch reported by \citet{Zhou2022} (UT 2022 March 28: sep = 0.600$''$~$\pm$~0.022$''$, PA = 182.5~$\pm$~1.4$\degr$), the \SNR\ = 0.8. We therefore do not detect emission in \PaBfilter\ from AB~Aur~b. At a separation of 0.6$''$, the achieved contrast $\Delta$\emph{Pa}$\beta$ for a spatially resolved source is \DeltaPaBetaExtended\ (\DeltaPaBeta\ for an unresolved source). Figure \ref{fig:2panel_5sigma_injected} shows four synthetic sources (both resolved and unresolved), injected into the image with fluxes equal to their respective 5$\sigma$ confidence level with separations 0.4$''$, 0.6$''$, 0.8$''$ and 1.0$''$ at PA = 270$\degr$. Each synthetic source is readily visible, and no other comparably bright features are evident elsewhere in the image. Although the continuum contrast between AB Aur b and its host star is lower, $\Delta J$ $\approx$ 9.2~mag \citep{Currie2022}, the host star is accreting and exhibits strong emission in \PaBfilter\, which will impact the contrast in this narrow band filter of the companion, and may explain why the companion was not recovered.

We also note an apparent region of elevated intensity near the expected location of AB~Aur~b (Figure \ref{fig:2panel}). The region appears to be spatially extended with peak intensity occurring at PA and separation of 184.0$\degr$ and 0.55$''$. The PA is similar to the nominal location of AB~Aur~b, yet its separation is smaller than the expected separation of the protoplanet. The peak \SNR\ in the region is 4.0, which remains below the 5$\sigma$ detection threshold. Deeper \PaBfilter\ observations could help clarify whether this is extended source is real and associated with AB~Aur~b.

\subsection{Monochromatic \PaBfilter\ Flux Density of AB~Aur~b}\label{subsec:Monochromatic_Flux_Density_Steps}

We compute an upper limit on the monochromatic flux density of AB~Aur~b from the upper limit of the apparent magnitude in \PaBfilter. Typically, this is determined using the apparent magnitude of the central star to calibrate the 5$\sigma$ contrast achieved at the expected separation of the companion:
\begin{equation}\label{eq:PaBetab}
    Pa\beta_\textrm{b} = \Delta Pa\beta + Pa\beta_\textrm{A}
\end{equation}
where $\Delta$\PaBfilter\ is the contrast at the separation of the companion, and \emph{Pa}$\beta_{A}$ is the apparent magnitude of the host star in \PaBfilter. AB Aur has a strong \PaBmag\ emission line, which may vary over time, so the apparent magnitude of the host star in that bandpass during the observations is not known.  To recover \emph{Pa}$\beta_{A}$ during the long-exposure images used in the high-contrast imaging sequence, we utilize the following relation,
\begin{equation}\label{eq:PaBetaA}
    Pa\beta_\textrm{A} = J_{cont,syn} - (J_{cont} - Pa\beta).
\end{equation}
Here, \Jcontsyn\ is a synthetic \Jcontfilter\ magnitude of AB~Aur~A, which we derived from our flux-calibrated IRTF spectrum\footnote{All synthetic photometry in this work was performed with the \texttt{pysynphot} tool set (\citealt{Lim2013}; \url{https://pysynphot.readthedocs.io/en/latest/}).}, and the (\JcontPaBcolorLabel) color is a single quantity denoting the stellar \JcontPaBcolorLabel\ color measured from the short-exposure (unsaturated) NIRC2 image sequences taken close together in time (within a 10 min time frame) immediately prior to the long-exposure ADI sequence. This method for obtaining \emph{Pa}$\beta_{A}$ assumes that AB~Aur~A's continuum in $J$ is constant. Long-term NIR monitoring by \citet{Shenavrin2019} shows that the \emph{J} magnitude of the central star varies by 0.37~mag on timescales of $\approx$1000 days (likely caused by dynamic processes in the inner region of the circumstellar disk; \citealt{Shenavrin2019}). Furthermore, accretion onto the star could fluctuate, which would vary the emission line strength in the \emph{J}-band region. For robustness, we adopt this uncertainty into our result for \Jcontsyn.


The IRTF spectrum is flux calibrated by matching the synthetic $J_{MKO}$ filter magnitude to 2MASS \emph{J}-band photometry (Figure \ref{fig:Spectrum}). We then performed synthetic photometry using the IRTF spectrum and the NIRC2 \Jcontfilter\ filter profile, returning \Jcontsyn\ = \Jcontsynthval.

Next, we calculated the \JcontPaBcolorLabel\ color of AB~Aur~A with the unsaturated \Jcontfilter\ and \PaBfilter\ images. For both filters, the detector counts were summed within a circular aperture centered on the stellar PSF, then divided by the integration times to give count rates in DN~s$^{-1}$. The aperture radius for \Jcontfilter\ was 2$\times$FWHM, and the radius for \PaBfilter\ was scaled according to the ratio of the central wavelengths of the \Jcontfilter\ and \PaBfilter\ filters\footnote{While this takes into account the slight difference in aperture radius between the \Jcontfilter\ ($\lambda_{0}$ = 1.2132) and \PaBfilter\ ($\lambda_{0}$ = 1.2903) filters, so as to preserve the same fractional energy in the PSF, we have assumed a constant Strehl ratio in both filters.}. The uncertainty of the total count rate is the standard deviation of that measured for each frame in the associated sequence. This procedure was repeated for the NIRC2 \Jcontfilter\ and \PaBfilter\ images of the A0 star HD~109691 from UT 2023 May 3, which was used to calibrate AB~Aur's \JcontPaBcolorLabel\ color to the Vega system as follows:
\begin{equation}\label{eq:Jcont-PaBeta}
    J_{cont} - Pa\beta = -2.5\log_{10}\frac{C_{Jcont}/C_{Pa\beta}}{C^{\textrm{A0}}_{Jcont}/C^{\textrm{A0}}_{Pa\beta}}
\end{equation}
where $C_{Jcont}$ and $C_{Pa\beta}$ are the measured detector count rates of AB Aur in each filter and $C^{\textrm{A0}}_{Jcont}$ and $C^{\textrm{A0}}_{Pa\beta}$ are the count rates measured for an A0 star in each filter. The count ratio $C^{\textrm{A0}}_{Jcont}/C^{\textrm{A0}}_{Pa\beta}$ is effectively the color of Vega (or any other A0V star) modulated by wavelength-dependent throughput losses from the atmosphere, telescope, filters, and detector. Using Equation \ref{eq:Jcont-PaBeta}, we find a color of \JcontPaBcolorLabel\ =~\JcontPaBetaColor.

\begin{figure}[!ht]
    \centering
    \includegraphics[width=\columnwidth]{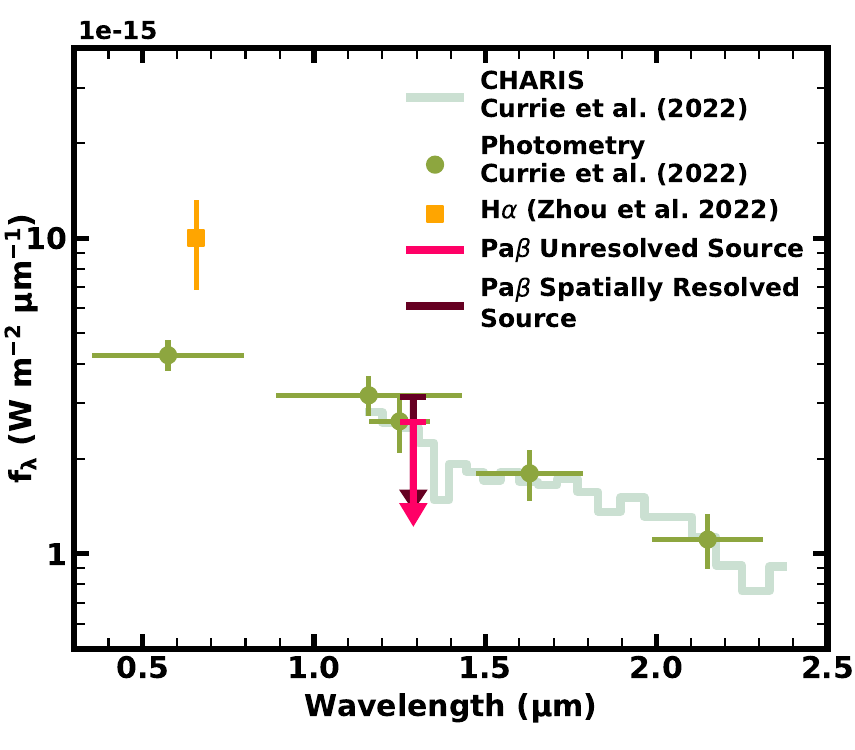}
    \caption{SED of AB~Aur~b. The adopted upper limit of the flux density in \PaBfilter\ for an unresolved source is indicated by the pink downward-pointing arrow, whereas the upper limit of the flux density for a spatially resolved source is shown depicted as a maroon arrow. The green circles are the optical (HST) and NIR (Subaru) photometry reported by \citet{Currie2022}, and the CHARIS spectrum they report is represented in light-green. The orange square point indicates the H$\alpha$ excess identified by \citet{Zhou2022} with HST.}
    \label{fig:SED}
\end{figure}

Using our results for $J_{cont,syn}$ and \JcontPaBcolorLabel, the \PaBfilter\ apparent magnitude of the star is \emph{Pa}$\beta_{A}$ = \PaBetaA\ at the time of our observations. In conjunction with the achieved contrast at 0.6$''$ separation ($\Delta$\PaBfilter\ = \DeltaPaBetaExtended\ for the resolved case, $\Delta$\PaBfilter\ = \DeltaPaBeta\ for the unresolved case; Section \ref{subsec:Sensitivity}), the 5$\sigma$ lower limit on the apparent magnitude of a spatially resolved AB~Aur~b is \emph{Pa}$\beta_{b}$ $>$ \PaBetabExtended\ (\PaBetab\ for the unresolved scenario). Applying a synthetic Vega \PaBfilter\ zero-point $f^{\textrm{Vega}}_{Pa\beta}$ = 2.50$\times$10$^{-9}$~W~m$^{-2}$~$\mu$m$^{-1}$, which was obtained with the Vega spectrum built into \texttt{pysynphot}\footnote{\Vegaurl}, the associated monochromatic flux density is $f_{Pa\beta}\leq$~\FluxDensityMeanExtended\ (\FluxDensityMean, unresolved). We proceed with a conservative treatment of $f_{Pa\beta}$, and adopt a hard upper limit equal to 2$\sigma$ above the mean value: henceforth, $f_{Pa\beta} \leq$ \FluxDensityLimitExtended\ (\FluxDensityLimit, unresolved; Figure \ref{fig:SED}).

\subsection{Pa$\beta$ Equivalent Width}\label{sec:EW}

\begin{figure}[!ht]
    \centering
    \includegraphics[width=\columnwidth]{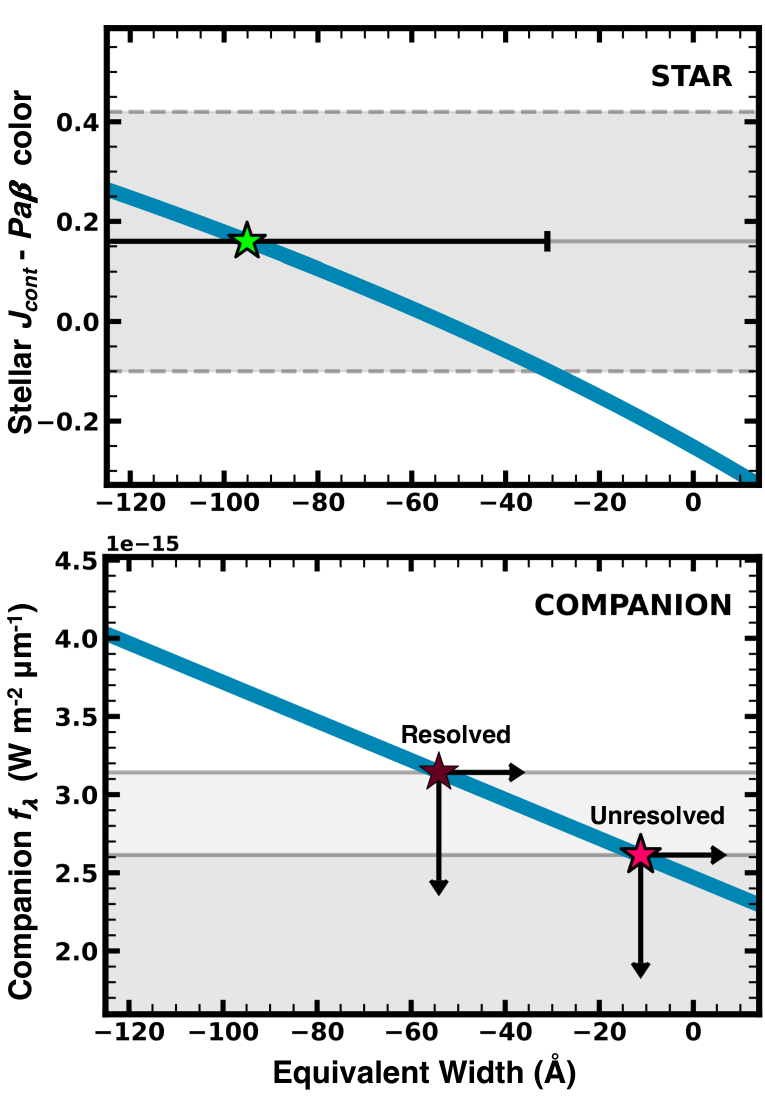}
    \caption{EWs and stellar color and companion flux density corresponding to simulated \PaBmag\ emission lines injected into the stellar continuum (top) and companion continuum (bottom). The solid grey line in the top panel indicates the real measured stellar \JcontPaBcolorLabel\ and the shaded region bounded by dashed lines represents the 1$\sigma$ uncertainty on the measurement. The green star marks the simulated quantity that matches the observed stellar \JcontPaBcolorLabel. The bottom panel shows the EW estimations for both an unresolved source and spatially resolved source. The uppermost boundary of each of the shaded regions indicates the upper limit of the companion flux density determined in Section \ref{subsec:Monochromatic_Flux_Density_Steps}. The pink and maroon stars mark the simulated quantities that match the measured upper limits on the companion's flux density if it were unresolved or resolved, respectively. EW values are negative for emission lines.}
    \label{fig:EW}
\end{figure}

To estimate the central star's EW at the time of imaging, we masked the \PaBmag\ emission line flux in the IRTF/SpeX SXD spectrum, and replaced the missing flux with a series of artificial emission lines with amplitudes ranging from --1 to 14 times the flux of the continuum at the line center, increasing sequentially by factors of 0.1. Each line is assigned a width corresponding to the SXD resolving power of $\approx$0.0006~$\mu$m. Next, the EW is calculated individually for each line and then mapped to the synthetic stellar \JcontPaBcolorLabel\ derived from the IRTF spectrum containing the corresponding simulated emission line. The true EW of the stellar \PaBmag\ emission line is then inferred from the simulated \JcontPaBcolorLabel\ color that matches the observed value (\JcontPaBcolorLabel~=~\JcontPaBetaColor; Figure \ref{fig:EW}). The results of this test indicate that the EW of the stellar \PaBmag\ line at the time of the imaging sequence is \EWstar. The EW of the \PaBmag\ emission line in the IRTF spectrum obtained $\sim$8 days later is $-11.58\,\pm\,0.09$~\AA~(Table \ref{tab:emission_lines}).

We obtained the companion's maximum detectable \PaBmag\ emission line strength with similar methods. We injected synthetic emission line profiles are into the CHARIS near-infrared spectrum of AB~Aur~b from \citet{Currie2022} following the same prescription for line amplitude and width as was done for the central star. The EW of each simulated emission line was mapped to the corresponding flux density of the continuum-plus-emission line through the NIRC2 \PaBfilter\ filter. Using this empirical relationship between \PaBfilter\ flux density and EW (Fig \ref{fig:EW}), we determine that AB~Aur~b's flux density upper limit $f_{\textrm{Pa}\beta} \leq$ \FluxDensityLimitExtended\ (\FluxDensityLimit, unresolved) corresponds to a lower\footnote{Noting that the EW for emission lines is negative, so weaker emission would yield higher values that are lower in magnitude.} limit in its EW of \EWcompanionExtended\ (\EWcompanion, unresolved). A significant difference in strength of the \PaBmag\ emission line between the star and the companion would be evidence that the companion is self-luminous (i.e., a protoplanet). However, their EWs agree at the 0.64$\sigma$ and 1.3$\sigma$ level for the spatially resolved and unresolved cases, respectively, offering no evidence of a significant ($\geq$3$\sigma$) discrepancy in \PaBfilter\ emission. This is consistent with both the light-scattering scenario as well as the scenario in which a young planet is weakly accreting.

\section{Discussion } \label{sec:Discussion}

Our deep \PaBfilter\ imaging does not reveal excess emission from AB~Aur~b, nor any other point source in the system. There is some tension between our upper limits at the location of the planet and the nominal published \emph{J}-band continuum level of AB~Aur~b. As described in Section \ref{subsec:Monochromatic_Flux_Density_Steps}, the 5$\sigma$ contrast level corresponds to an upper limit of \FluxDensityMeanExtended\ assuming a spatially resolved source (or \FluxDensityMean\ for the unresolved case in which AB~Aur~b is a point source), which is below the continuum emission reported by \citet{Currie2022} at this wavelength. But as there is an uncertainty associated with our upper limit, we have conservatively adopted the 95\% upper bound on this value to guard against potential systematic errors in the multi-step process of converting contrast measurements into a flux density within this narrow-band filter.

There are a few potential explanations that can account for the companion flux being undetectable at the published continuum level:
\begin{itemize}
    \item If we consider the companion to be a self-luminous protoplanet, we can rule out strong \PaBfilter\ emission, implying weak accretion or no accretion at all from the companion.

    \item Alternatively, if the companion's emission source is a compact protoplanetary disk feature seen in scattered starlight, one or both of the following may be possible. The localized dynamics in the disk material could have changed the effective light-scattering area of the disk substructure (e.g., LkCa 15 \citealt{Thalmann2016, Currie2019, Sallum2023}). Alternatively, there could have been enhanced \PaBfilter\ emission caused by accretion onto the central star, but the delay in light travel time to reach the companion may have prevented the observation of enhanced emission at the location of AB~Aur~b within the observing window.

    \item Finally, irrespective of the true nature of the emission source, the following two possibilities could also contribute to this modest tension: the \emph{J}-band region is known to be moderately variable at the 0.4~mag level \citep{Shenavrin2019}, and could have affected the continuum flux calibration (see also Section \ref{subsec:Monochromatic_Flux_Density_Steps}).
    
    \item Alternatively, the absolute calibration level of the CHARIS continuum may be inaccurate. For this analysis, we have assumed that the companion's continuum is both constant and accurate.

\end{itemize}

\section{Summary} \label{sec:summary}

In this work, we obtained high-contrast imaging of AB Aur with Keck/NIRC2 in the \PaBfilter\ narrow-band filter. We do not detect significant \PaBfilter\ emission in the FOV surrounding the star, either from AB Aur b or any other potential embedded planets. However, the depth of our observations place a meaningful upper limit on the \PaBmag\ line flux. Mapping this flux to its associated EW, we find that, if present, \PaBfilter\ emission is weak. However, we cannot conclude whether it must be significantly weaker in comparison to emission from the central star at the time of the observations, which would have provided a direct test of the scattered light versus self-luminous hypothesis because the line-to-continuum ratio is expected to differ in the latter scenario. Deeper \PaBfilter\ imaging, NIR spectroscopy, and multi-wavelength observations of AB~Aur~b will help clarify the nature of this enigmatic source.

\section*{Acknowledgements}
We thank the staff at Keck Observatory for their assistance in enabling these observations and their execution. B.P.B. acknowledges support from the National Science Foundation grant AST-1909209, NASA Exoplanet Research Program grant 20-XRP20$\_$2-0119, and the Alfred P. Sloan Foundation. This work made use of the following software: \texttt{Astropy} \citep{Robitaille2021}, \texttt{Numpy} \citep{vanderWalt2011}, \texttt{Scipy} \citep{Virtanen2021}, \texttt{Matplotlib} \citep{Hunter2007}, \texttt{pyKLIP} \citep{Wang2015}, and \texttt{pysynphot} \citep{Lim2013}.

\bibliography{Main_v6.bib}{}

\begin{thebibliography}{}
\expandafter\ifx\csname natexlab\endcsname\relax\def\natexlab#1{#1}\fi
\providecommand{\url}[1]{\href{#1}{#1}}
\providecommand{\dodoi}[1]{doi:~\href{http://doi.org/#1}{\nolinkurl{#1}}}
\providecommand{\doeprint}[1]{\href{http://ascl.net/#1}{\nolinkurl{http://ascl.net/#1}}}
\providecommand{\doarXiv}[1]{\href{https://arxiv.org/abs/#1}{\nolinkurl{https://arxiv.org/abs/#1}}}

\bibitem[{{Adams Redai} {et~al.}(2023){Adams Redai}, {Follette}, {Wang},
  {Leonard}, {Balmer}, {Close}, {Dacus}, {Males}, {Morzinski}, {Palmo},
  {Pueyo}, {Spiro}, {Treiber}, {Ward-Duong}, \& {Watson}}]{Adams2023}
{Adams Redai}, J.~I., {Follette}, K.~B., {Wang}, J., {et~al.} 2023, \aj, 165,
  57, \dodoi{10.3847/1538-3881/aca60d}

\bibitem[{{Aoyama} {et~al.}(2018){Aoyama}, {Ikoma}, \& {Tanigawa}}]{Aoyama2018}
{Aoyama}, Y., {Ikoma}, M., \& {Tanigawa}, T. 2018, \apj, 866, 84,
  \dodoi{10.3847/1538-4357/aadc11}

\bibitem[{{Aoyama} {et~al.}(2020){Aoyama}, {Marleau}, {Mordasini}, \&
  {Ikoma}}]{Aoyama2020}
{Aoyama}, Y., {Marleau}, G.-D., {Mordasini}, C., \& {Ikoma}, M. 2020, arXiv
  e-prints, arXiv:2011.06608, \dodoi{10.48550/arXiv.2011.06608}

\bibitem[{{Biller} {et~al.}(2014){Biller}, {Males}, {Rodigas}, {Morzinski},
  {Close}, {Juh{\'a}sz}, {Follette}, {Lacour}, {Benisty}, {Sicilia-Aguilar},
  {Hinz}, {Weinberger}, {Henning}, {Pott}, {Bonnefoy}, \&
  {K{\"o}hler}}]{Biller2014}
{Biller}, B.~A., {Males}, J., {Rodigas}, T., {et~al.} 2014, \apjl, 792, L22,
  \dodoi{10.1088/2041-8205/792/1/L22}

\bibitem[{{Bowler} {et~al.}(2011){Bowler}, {Liu}, {Kraus}, {Mann}, \&
  {Ireland}}]{Bowler2011}
{Bowler}, B.~P., {Liu}, M.~C., {Kraus}, A.~L., {Mann}, A.~W., \& {Ireland},
  M.~J. 2011, \apj, 743, 148, \dodoi{10.1088/0004-637X/743/2/148}

\bibitem[{{Choksi} \& {Chiang}(2022)}]{Choksi2022}
{Choksi}, N., \& {Chiang}, E. 2022, \mnras, 510, 1657,
  \dodoi{10.1093/mnras/stab3503}

\bibitem[{{Costigan} {et~al.}(2014){Costigan}, {Vink}, {Scholz}, {Ray}, \&
  {Testi}}]{Costigan2014}
{Costigan}, G., {Vink}, J.~S., {Scholz}, A., {Ray}, T., \& {Testi}, L. 2014,
  \mnras, 440, 3444, \dodoi{10.1093/mnras/stu529}

\bibitem[{{Currie} {et~al.}(2019){Currie}, {Marois}, {Cieza}, {Mulders},
  {Lawson}, {Caceres}, {Rodriguez-Ruiz}, {Wisniewski}, {Guyon}, {Brandt},
  {Kasdin}, {Groff}, {Lozi}, {Chilcote}, {Hodapp}, {Jovanovic}, {Martinache},
  {Skaf}, {Lyra}, {Tamura}, {Asensio-Torres}, {Dong}, {Grady}, {Gerard},
  {Fukagawa}, {Hand}, {Hayashi}, {Henning}, {Kudo}, {Kuzuhara}, {Kwon},
  {McElwain}, \& {Uyama}}]{Currie2019}
{Currie}, T., {Marois}, C., {Cieza}, L., {et~al.} 2019, \apjl, 877, L3,
  \dodoi{10.3847/2041-8213/ab1b42}

\bibitem[{{Currie} {et~al.}(2022){Currie}, {Lawson}, {Schneider}, {Lyra},
  {Wisniewski}, {Grady}, {Guyon}, {Tamura}, {Kotani}, {Kawahara}, {Brandt},
  {Uyama}, {Muto}, {Dong}, {Kudo}, {Hashimoto}, {Fukagawa}, {Wagner}, {Lozi},
  {Chilcote}, {Tobin}, {Groff}, {Ward-Duong}, {Januszewski}, {Norris},
  {Tuthill}, {van der Marel}, {Sitko}, {Deo}, {Vievard}, {Jovanovic},
  {Martinache}, \& {Skaf}}]{Currie2022}
{Currie}, T., {Lawson}, K., {Schneider}, G., {et~al.} 2022, Nature Astronomy,
  6, 751, \dodoi{10.1038/s41550-022-01634-x}

\bibitem[{{Cushing} {et~al.}(2004){Cushing}, {Vacca}, \&
  {Rayner}}]{Cushing2004}
{Cushing}, M.~C., {Vacca}, W.~D., \& {Rayner}, J.~T. 2004, \pasp, 116, 362,
  \dodoi{10.1086/382907}

\bibitem[{{Demars} {et~al.}(2023){Demars}, {Bonnefoy}, {Dougados}, {Aoyama},
  {Thanathibodee}, {Marleau}, {Tremblin}, {Delorme}, {Palma-Bifani}, {Petrus},
  {Bowler}, {Chauvin}, \& {Lagrange}}]{Demars2023}
{Demars}, D., {Bonnefoy}, M., {Dougados}, C., {et~al.} 2023, arXiv e-prints,
  arXiv:2305.09460.
\newblock \doarXiv{2305.09460}

\bibitem[{{Follette} {et~al.}(2017){Follette}, {Rameau}, {Dong}, {Pueyo},
  {Close}, {Duch{\^e}ne}, {Fung}, {Leonard}, {Macintosh}, {Males}, {Marois},
  {Millar-Blanchaer}, {Morzinski}, {Mullen}, {Perrin}, {Spiro}, {Wang},
  {Ammons}, {Bailey}, {Barman}, {Bulger}, {Chilcote}, {Cotten}, {De Rosa},
  {Doyon}, {Fitzgerald}, {Goodsell}, {Graham}, {Greenbaum}, {Hibon}, {Hung},
  {Ingraham}, {Kalas}, {Konopacky}, {Larkin}, {Maire}, {Marchis}, {Metchev},
  {Nielsen}, {Oppenheimer}, {Palmer}, {Patience}, {Poyneer}, {Rajan},
  {Rantakyr{\"o}}, {Savransky}, {Schneider}, {Sivaramakrishnan}, {Song},
  {Soummer}, {Thomas}, {Vega}, {Wallace}, {Ward-Duong}, {Wiktorowicz}, \&
  {Wolff}}]{Follette2017}
{Follette}, K.~B., {Rameau}, J., {Dong}, R., {et~al.} 2017, \aj, 153, 264,
  \dodoi{10.3847/1538-3881/aa6d85}

\bibitem[{{Follette} {et~al.}(2022){Follette}, {Close}, {Males}, {Ward-Duong},
  {Balmer}, {Adams Redai}, {Morales}, {Sarosi}, {Dacus}, {De Rosa}, {Garcia
  Toro}, {Leonard}, {Macintosh}, {Morzinski}, {Mullen}, {Palmo}, {Nzaba
  Saitoti}, {Spiro}, {Treiber}, {Wang}, {Wang}, {Watson}, \&
  {Weinberger}}]{Follette2022}
{Follette}, K.~B., {Close}, L.~M., {Males}, J.~R., {et~al.} 2022, arXiv
  e-prints, arXiv:2211.02109, \dodoi{10.48550/arXiv.2211.02109}

\bibitem[{{Fruchter} \& {Hook}(2002)}]{FruchterHook2002}
{Fruchter}, A.~S., \& {Hook}, R.~N. 2002, \pasp, 114, 144,
  \dodoi{10.1086/338393}

\bibitem[{{Gaia Collaboration}(2020)}]{GaiaCollaboration2020}
{Gaia Collaboration}. 2020, VizieR Online Data Catalog, I/350

\bibitem[{{Guzm{\'a}n-D{\'\i}az} {et~al.}(2021){Guzm{\'a}n-D{\'\i}az},
  {Mendigut{\'\i}a}, {Montesinos}, {Oudmaijer}, {Vioque}, {Rodrigo}, {Solano},
  {Meeus}, \& {Marcos-Arenal}}]{GuzmanDiaz2021}
{Guzm{\'a}n-D{\'\i}az}, J., {Mendigut{\'\i}a}, I., {Montesinos}, B., {et~al.}
  2021, \aap, 650, A182, \dodoi{10.1051/0004-6361/202039519}

\bibitem[{{Haffert} {et~al.}(2019){Haffert}, {Bohn}, {de Boer}, {Snellen},
  {Brinchmann}, {Girard}, {Keller}, \& {Bacon}}]{Haffert2019}
{Haffert}, S.~Y., {Bohn}, A.~J., {de Boer}, J., {et~al.} 2019, Nature
  Astronomy, 3, 749, \dodoi{10.1038/s41550-019-0780-5}

\bibitem[{{Harrington} \& {Kuhn}(2007)}]{HarringtonKuhn2007}
{Harrington}, D.~M., \& {Kuhn}, J.~R. 2007, \apjl, 667, L89,
  \dodoi{10.1086/521999}

\bibitem[{Hunter(2007)}]{Hunter2007}
Hunter, J.~D. 2007, Computing in Science \& Engineering, 9, 90,
  \dodoi{10.1109/MCSE.2007.55}

\bibitem[{{Keppler} {et~al.}(2018){Keppler}, {Benisty}, {M{\"u}ller},
  {Henning}, {van Boekel}, {Cantalloube}, {Ginski}, {van Holstein}, {Maire},
  {Pohl}, {Samland}, {Avenhaus}, {Baudino}, {Boccaletti}, {de Boer},
  {Bonnefoy}, {Chauvin}, {Desidera}, {Langlois}, {Lazzoni}, {Marleau},
  {Mordasini}, {Pawellek}, {Stolker}, {Vigan}, {Zurlo}, {Birnstiel},
  {Brandner}, {Feldt}, {Flock}, {Girard}, {Gratton}, {Hagelberg}, {Isella},
  {Janson}, {Juhasz}, {Kemmer}, {Kral}, {Lagrange}, {Launhardt}, {Matter},
  {M{\'e}nard}, {Milli}, {Molli{\`e}re}, {Olofsson}, {P{\'e}rez}, {Pinilla},
  {Pinte}, {Quanz}, {Schmidt}, {Udry}, {Wahhaj}, {Williams}, {Buenzli},
  {Cudel}, {Dominik}, {Galicher}, {Kasper}, {Lannier}, {Mesa}, {Mouillet},
  {Peretti}, {Perrot}, {Salter}, {Sissa}, {Wildi}, {Abe}, {Antichi},
  {Augereau}, {Baruffolo}, {Baudoz}, {Bazzon}, {Beuzit}, {Blanchard}, {Brems},
  {Buey}, {De Caprio}, {Carbillet}, {Carle}, {Cascone}, {Cheetham}, {Claudi},
  {Costille}, {Delboulb{\'e}}, {Dohlen}, {Fantinel}, {Feautrier}, {Fusco},
  {Giro}, {Gluck}, {Gry}, {Hubin}, {Hugot}, {Jaquet}, {Le Mignant}, {Llored},
  {Madec}, {Magnard}, {Martinez}, {Maurel}, {Meyer}, {M{\"o}ller-Nilsson},
  {Moulin}, {Mugnier}, {Orign{\'e}}, {Pavlov}, {Perret}, {Petit}, {Pragt},
  {Puget}, {Rabou}, {Ramos}, {Rigal}, {Rochat}, {Roelfsema}, {Rousset}, {Roux},
  {Salasnich}, {Sauvage}, {Sevin}, {Soenke}, {Stadler}, {Suarez}, {Turatto}, \&
  {Weber}}]{Keppler2018}
{Keppler}, M., {Benisty}, M., {M{\"u}ller}, A., {et~al.} 2018, \aap, 617, A44,
  \dodoi{10.1051/0004-6361/201832957}

\bibitem[{{Ligi} {et~al.}(2018){Ligi}, {Vigan}, {Gratton}, {de Boer},
  {Benisty}, {Boccaletti}, {Quanz}, {Meyer}, {Ginski}, {Sissa}, {Gry},
  {Henning}, {Beuzit}, {Biller}, {Bonnefoy}, {Chauvin}, {Cheetham}, {Cudel},
  {Delorme}, {Desidera}, {Feldt}, {Galicher}, {Girard}, {Janson}, {Kasper},
  {Kopytova}, {Lagrange}, {Langlois}, {Lecoroller}, {Maire}, {M{\'e}nard},
  {Mesa}, {Peretti}, {Perrot}, {Pinilla}, {Pohl}, {Rouan}, {Stolker},
  {Samland}, {Wahhaj}, {Wildi}, {Zurlo}, {Buey}, {Fantinel}, {Fusco}, {Jaquet},
  {Moulin}, {Ramos}, {Suarez}, \& {Weber}}]{Ligi2018}
{Ligi}, R., {Vigan}, A., {Gratton}, R., {et~al.} 2018, \mnras, 473, 1774,
  \dodoi{10.1093/mnras/stx2318}

\bibitem[{{Lim} {et~al.}(2013){Lim}, {Diaz}, \& {Laidler}}]{Lim2013}
{Lim}, P.~L., {Diaz}, R.~I., \& {Laidler}, V. 2013, {pysynphot: Synthetic
  photometry software package}, Astrophysics Source Code Library, record
  ascl:1303.023.
\newblock \doeprint{1303.023}

\bibitem[{{Liu}(2004)}]{Liu2004}
{Liu}, M.~C. 2004, Science, 305, 1442, \dodoi{10.1126/science.1102929}

\bibitem[{{Marleau} \& {Aoyama}(2022)}]{Marleau2022}
{Marleau}, G.-D., \& {Aoyama}, Y. 2022, Research Notes of the American
  Astronomical Society, 6, 262, \dodoi{10.3847/2515-5172/acaa34}

\bibitem[{{Marois} {et~al.}(2006){Marois}, {Lafreni{\`e}re}, {Doyon},
  {Macintosh}, \& {Nadeau}}]{Marois2006}
{Marois}, C., {Lafreni{\`e}re}, D., {Doyon}, R., {Macintosh}, B., \& {Nadeau},
  D. 2006, \apj, 641, 556, \dodoi{10.1086/500401}

\bibitem[{{Mawet} {et~al.}(2014){Mawet}, {Milli}, {Wahhaj}, {Pelat}, {Absil},
  {Delacroix}, {Boccaletti}, {Kasper}, {Kenworthy}, {Marois}, {Mennesson}, \&
  {Pueyo}}]{Mawet2014}
{Mawet}, D., {Milli}, J., {Wahhaj}, Z., {et~al.} 2014, \apj, 792, 97,
  \dodoi{10.1088/0004-637X/792/2/97}

\bibitem[{{Mendigut{\'\i}a} {et~al.}(2018){Mendigut{\'\i}a}, {Oudmaijer},
  {Schneider}, {Hu{\'e}lamo}, {Baines}, {Brittain}, \&
  {Aberasturi}}]{Mendigutia2018}
{Mendigut{\'\i}a}, I., {Oudmaijer}, R.~D., {Schneider}, P.~C., {et~al.} 2018,
  \aap, 618, L9, \dodoi{10.1051/0004-6361/201834233}

\bibitem[{{Pohl} {et~al.}(2017){Pohl}, {Benisty}, {Pinilla}, {Ginski}, {de
  Boer}, {Avenhaus}, {Henning}, {Zurlo}, {Boccaletti}, {Augereau}, {Birnstiel},
  {Dominik}, {Facchini}, {Fedele}, {Janson}, {Keppler}, {Kral}, {Langlois},
  {Ligi}, {Maire}, {M{\'e}nard}, {Meyer}, {Pinte}, {Quanz}, {Sauvage},
  {Sezestre}, {Stolker}, {Szul{\'a}gyi}, {van Boekel}, {van der Plas},
  {Villenave}, {Baruffolo}, {Baudoz}, {Le Mignant}, {Maurel}, {Ramos}, \&
  {Weber}}]{Pohl2017}
{Pohl}, A., {Benisty}, M., {Pinilla}, P., {et~al.} 2017, \apj, 850, 52,
  \dodoi{10.3847/1538-4357/aa94c2}

\bibitem[{{Rameau} {et~al.}(2017){Rameau}, {Follette}, {Pueyo}, {Marois},
  {Macintosh}, {Millar-Blanchaer}, {Wang}, {Vega}, {Doyon}, {Lafreni{\`e}re},
  {Nielsen}, {Bailey}, {Chilcote}, {Close}, {Esposito}, {Males}, {Metchev},
  {Morzinski}, {Ruffio}, {Wolff}, {Ammons}, {Barman}, {Bulger}, {Cotten}, {De
  Rosa}, {Duchene}, {Fitzgerald}, {Goodsell}, {Graham}, {Greenbaum}, {Hibon},
  {Hung}, {Ingraham}, {Kalas}, {Konopacky}, {Larkin}, {Maire}, {Marchis},
  {Oppenheimer}, {Palmer}, {Patience}, {Perrin}, {Poyneer}, {Rajan},
  {Rantakyr{\"o}}, {Marley}, {Savransky}, {Schneider}, {Sivaramakrishnan},
  {Song}, {Soummer}, {Thomas}, {Wallace}, {Ward-Duong}, \&
  {Wiktorowicz}}]{Rameau2017}
{Rameau}, J., {Follette}, K.~B., {Pueyo}, L., {et~al.} 2017, \aj, 153, 244,
  \dodoi{10.3847/1538-3881/aa6cae}

\bibitem[{{Rayner} {et~al.}(2003){Rayner}, {Toomey}, {Onaka}, {Denault},
  {Stahlberger}, {Vacca}, {Cushing}, \& {Wang}}]{Rayner2003}
{Rayner}, J.~T., {Toomey}, D.~W., {Onaka}, P.~M., {et~al.} 2003, \pasp, 115,
  362, \dodoi{10.1086/367745}

\bibitem[{{Robitaille} {et~al.}(2021){Robitaille}, {Tollerud}, {Aldcroft},
  {Bray}, {Van Kerkwijk}, {Droettboom}, {Sip{\H{o}}cz}, {Lim}, {Price-Whelan},
  {Conseil}, {Dencheva}, {Bradley}, {Ginsburg}, {Mumford}, {Seifert}, {Craig},
  {Mdmueller}, {StuartLittlefair}, {D'Avella}, {Lglattly}, {G{\"u}nther},
  {Homeier}, {Donath}, {Perrygreenfield}, {Deil}, {Vin{\'\i}cius}, {Woillez},
  {Vanderplas}, {Patil}, \& {N{\"o}the}}]{Robitaille2021}
{Robitaille}, T., {Tollerud}, E., {Aldcroft}, T., {et~al.} 2021,
  {astropy/astropy: v4.2.1}, v4.2.1, Zenodo,  Zenodo,
  \dodoi{10.5281/zenodo.4670729}

\bibitem[{{Sallum} {et~al.}(2023){Sallum}, {Eisner}, {Skemer}, \&
  {Murray-Clay}}]{Sallum2023}
{Sallum}, S., {Eisner}, J., {Skemer}, A., \& {Murray-Clay}, R. 2023, \apj, 953,
  55, \dodoi{10.3847/1538-4357/ace16c}

\bibitem[{{Santamar{\'\i}a-Miranda} {et~al.}(2018){Santamar{\'\i}a-Miranda},
  {C{\'a}ceres}, {Schreiber}, {Hardy}, {Bayo}, {Parsons}, {Gromadzki}, \&
  {Aguayo Villegas}}]{Santamaria-Miranda2018}
{Santamar{\'\i}a-Miranda}, A., {C{\'a}ceres}, C., {Schreiber}, M.~R., {et~al.}
  2018, \mnras, 475, 2994, \dodoi{10.1093/mnras/stx3325}

\bibitem[{{Schmidt} {et~al.}(2008){Schmidt}, {Neuh{\"a}user}, {Seifahrt},
  {Vogt}, {Bedalov}, {Helling}, {Witte}, \& {Hauschildt}}]{Schmidt2008}
{Schmidt}, T.~O.~B., {Neuh{\"a}user}, R., {Seifahrt}, A., {et~al.} 2008, \aap,
  491, 311, \dodoi{10.1051/0004-6361:20078840}

\bibitem[{{Seifahrt} {et~al.}(2007){Seifahrt}, {Neuh{\"a}user}, \&
  {Hauschildt}}]{Seifahrt2007}
{Seifahrt}, A., {Neuh{\"a}user}, R., \& {Hauschildt}, P.~H. 2007, \aap, 463,
  309, \dodoi{10.1051/0004-6361:20066463}

\bibitem[{{Service} {et~al.}(2016){Service}, {Lu}, {Campbell}, {Sitarski},
  {Ghez}, \& {Anderson}}]{Service2016}
{Service}, M., {Lu}, J.~R., {Campbell}, R., {et~al.} 2016, \pasp, 128, 095004,
  \dodoi{10.1088/1538-3873/128/967/095004}

\bibitem[{{Shenavrin} {et~al.}(2019){Shenavrin}, {Grinin}, {Baluev}, \&
  {Demidova}}]{Shenavrin2019}
{Shenavrin}, V.~I., {Grinin}, V.~P., {Baluev}, R.~V., \& {Demidova}, T.~V.
  2019, Astronomy Reports, 63, 1035, \dodoi{10.1134/S1063772919120060}

\bibitem[{{Sissa} {et~al.}(2018){Sissa}, {Gratton}, {Garufi}, {Rigliaco},
  {Zurlo}, {Mesa}, {Langlois}, {de Boer}, {Desidera}, {Ginski}, {Lagrange},
  {Maire}, {Vigan}, {Dima}, {Antichi}, {Baruffolo}, {Bazzon}, {Benisty},
  {Beuzit}, {Biller}, {Boccaletti}, {Bonavita}, {Bonnefoy}, {Brandner},
  {Bruno}, {Buenzli}, {Cascone}, {Chauvin}, {Cheetham}, {Claudi}, {Cudel}, {De
  Caprio}, {Dominik}, {Fantinel}, {Farisato}, {Feldt}, {Fontanive}, {Galicher},
  {Giro}, {Hagelberg}, {Incorvaia}, {Janson}, {Kasper}, {Keppler}, {Kopytova},
  {Lagadec}, {Lannier}, {Lazzoni}, {LeCoroller}, {Lessio}, {Ligi}, {Marzari},
  {Menard}, {Meyer}, {Mouillet}, {Peretti}, {Perrot}, {Potiron}, {Rouan},
  {Salasnich}, {Salter}, {Samland}, {Schmidt}, {Scuderi}, \&
  {Wildi}}]{Sissa2018}
{Sissa}, E., {Gratton}, R., {Garufi}, A., {et~al.} 2018, \aap, 619, A160,
  \dodoi{10.1051/0004-6361/201732332}

\bibitem[{{Stolker} {et~al.}(2021){Stolker}, {Haffert}, {Kesseli}, {van
  Holstein}, {Aoyama}, {Brinchmann}, {Cugno}, {Girard}, {Marleau}, {Meyer},
  {Milli}, {Quanz}, {Snellen}, \& {Todorov}}]{Stolker2021}
{Stolker}, T., {Haffert}, S.~Y., {Kesseli}, A.~Y., {et~al.} 2021, \aj, 162,
  286, \dodoi{10.3847/1538-3881/ac2c7f}

\bibitem[{{Szul{\'a}gyi} \& {Ercolano}(2020)}]{Szulagyi2020}
{Szul{\'a}gyi}, J., \& {Ercolano}, B. 2020, \apj, 902, 126,
  \dodoi{10.3847/1538-4357/abb5a2}

\bibitem[{{Szul{\'a}gyi} {et~al.}(2014){Szul{\'a}gyi}, {Morbidelli}, {Crida},
  \& {Masset}}]{Szulagyi2014}
{Szul{\'a}gyi}, J., {Morbidelli}, A., {Crida}, A., \& {Masset}, F. 2014, \apj,
  782, 65, \dodoi{10.1088/0004-637X/782/2/65}

\bibitem[{{Tang} {et~al.}(2017){Tang}, {Guilloteau}, {Dutrey}, {Muto}, {Shen},
  {Gu}, {Inutsuka}, {Momose}, {Pietu}, {Fukagawa}, {Chapillon}, {Ho}, {di
  Folco}, {Corder}, {Ohashi}, \& {Hashimoto}}]{Tang2017}
{Tang}, Y.-W., {Guilloteau}, S., {Dutrey}, A., {et~al.} 2017, \apj, 840, 32,
  \dodoi{10.3847/1538-4357/aa6af7}

\bibitem[{{Tanigawa} {et~al.}(2012){Tanigawa}, {Ohtsuki}, \&
  {Machida}}]{Tanigawa2012}
{Tanigawa}, T., {Ohtsuki}, K., \& {Machida}, M.~N. 2012, \apj, 747, 47,
  \dodoi{10.1088/0004-637X/747/1/47}

\bibitem[{{Thalmann} {et~al.}(2016){Thalmann}, {Janson}, {Garufi},
  {Boccaletti}, {Quanz}, {Sissa}, {Gratton}, {Salter}, {Benisty}, {Bonnefoy},
  {Chauvin}, {Daemgen}, {Desidera}, {Dominik}, {Engler}, {Feldt}, {Henning},
  {Lagrange}, {Langlois}, {Lannier}, {Le Coroller}, {Ligi}, {M{\'e}nard},
  {Mesa}, {Meyer}, {Mulders}, {Olofsson}, {Pinte}, {Schmid}, {Vigan}, \&
  {Zurlo}}]{Thalmann2016}
{Thalmann}, C., {Janson}, M., {Garufi}, A., {et~al.} 2016, \apjl, 828, L17,
  \dodoi{10.3847/2041-8205/828/2/L17}

\bibitem[{{Thanathibodee} {et~al.}(2019){Thanathibodee}, {Calvet}, {Bae},
  {Muzerolle}, \& {Hern{\'a}ndez}}]{Thanathibodee2019}
{Thanathibodee}, T., {Calvet}, N., {Bae}, J., {Muzerolle}, J., \&
  {Hern{\'a}ndez}, R.~F. 2019, \apj, 885, 94, \dodoi{10.3847/1538-4357/ab44c1}

\bibitem[{{van Holstein} {et~al.}(2021){van Holstein}, {Stolker},
  {Jensen-Clem}, {Ginski}, {Milli}, {de Boer}, {Girard}, {Wahhaj}, {Bohn},
  {Millar-Blanchaer}, {Benisty}, {Bonnefoy}, {Chauvin}, {Dominik}, {Hinkley},
  {Keller}, {Keppler}, {Langlois}, {Marino}, {M{\'e}nard}, {Perrot}, {Schmidt},
  {Vigan}, {Zurlo}, \& {Snik}}]{vanHolstein2021}
{van Holstein}, R.~G., {Stolker}, T., {Jensen-Clem}, R., {et~al.} 2021, \aap,
  647, A21, \dodoi{10.1051/0004-6361/202039290}

\bibitem[{vanderWalt2011 {et~al.}(2011)vanderWalt2011, Colbert, \&
  Varoquaux}]{vanderWalt2011}
vanderWalt2011, S., Colbert, S.~C., \& Varoquaux, G. 2011, Computing in Science
  \& Engineering, 13, 22, \dodoi{10.1109/MCSE.2011.37}

\bibitem[{{Virtanen} {et~al.}(2021){Virtanen}, {Gommers}, {Burovski},
  {Oliphant}, {Weckesser}, {Cournapeau}, {Alexbrc}, {Reddy}, {Peterson},
  {Haberland}, {Wilson}, {Nelson}, {Endolith}, {Mayorov}, {Van Der Walt},
  {Polat}, {Laxalde}, {Brett}, {Larson}, {Millman}, {Lars}, {Van Mulbregt},
  {Eric-Jones}, {Carey}, {Peterbell10}, {Moore}, {Kern}, {Leslie}, {Perktold},
  \& {Striega}}]{Virtanen2021}
{Virtanen}, P., {Gommers}, R., {Burovski}, E., {et~al.} 2021, {scipy/scipy:
  SciPy 1.6.3}, v1.6.3, Zenodo,  Zenodo, \dodoi{10.5281/zenodo.4718897}

\bibitem[{{Wagner} {et~al.}(2018){Wagner}, {Follete}, {Close}, {Apai}, {Gibbs},
  {Keppler}, {M{\"u}ller}, {Henning}, {Kasper}, {Wu}, {Long}, {Males},
  {Morzinski}, \& {McClure}}]{Wagner2018}
{Wagner}, K., {Follete}, K.~B., {Close}, L.~M., {et~al.} 2018, \apjl, 863, L8,
  \dodoi{10.3847/2041-8213/aad695}

\bibitem[{{Wang} {et~al.}(2015){Wang}, {Ruffio}, {De Rosa}, {Aguilar}, {Wolff},
  \& {Pueyo}}]{Wang2015}
{Wang}, J.~J., {Ruffio}, J.-B., {De Rosa}, R.~J., {et~al.} 2015, {pyKLIP: PSF
  Subtraction for Exoplanets and Disks}, Astrophysics Source Code Library,
  record ascl:1506.001.
\newblock \doeprint{1506.001}

\bibitem[{{Wizinowich} {et~al.}(2000){Wizinowich}, {Acton}, {Shelton},
  {Stomski}, {Gathright}, {Ho}, {Lupton}, {Tsubota}, {Lai}, {Max}, {Brase},
  {An}, {Avicola}, {Olivier}, {Gavel}, {Macintosh}, {Ghez}, \&
  {Larkin}}]{Wizinowich2000}
{Wizinowich}, P., {Acton}, D.~S., {Shelton}, C., {et~al.} 2000, \pasp, 112,
  315, \dodoi{10.1086/316543}

\bibitem[{{Zhou} {et~al.}(2014){Zhou}, {Herczeg}, {Kraus}, {Metchev}, \&
  {Cruz}}]{Zhou2014}
{Zhou}, Y., {Herczeg}, G.~J., {Kraus}, A.~L., {Metchev}, S., \& {Cruz}, K.~L.
  2014, \apjl, 783, L17, \dodoi{10.1088/2041-8205/783/1/L17}

\bibitem[{{Zhou} {et~al.}(2022){Zhou}, {Sanghi}, {Bowler}, {Wu}, {Close},
  {Long}, {Ward-Duong}, {Zhu}, {Kraus}, {Follette}, \& {Bae}}]{Zhou2022}
{Zhou}, Y., {Sanghi}, A., {Bowler}, B.~P., {et~al.} 2022, \apjl, 934, L13,
  \dodoi{10.3847/2041-8213/ac7fef}

\bibitem[{{Zhou} {et~al.}(2023){Zhou}, {Bowler}, {Yang}, {Sanghi}, {Herczeg},
  {Kraus}, {Bae}, {Long}, {Follette}, {Ward-Duong}, {Zhu}, {Biddle}, {Close},
  {Yushu Jiang}, \& {Wu}}]{Zhou2023}
{Zhou}, Y., {Bowler}, B.~P., {Yang}, H., {et~al.} 2023, arXiv e-prints,
  arXiv:2308.16223, \dodoi{10.48550/arXiv.2308.16223}

\end{thebibliography}
\bibliographystyle{aasjournal}



\FloatBarrier

\appendix
\vspace{-12pt}
\section{Pa$\beta$ Contrast Curve}\label{appendix:contrast}

\begin{deluxetable}{ccc}[h!]
\tabletypesize{\normalsize}
\tablecolumns{3}
\tablewidth{0pt}
\tableheadfrac{0.2}
\tablecaption{Separation vs. \PaBfilter\ contrast (Figure \ref{fig:contrast_curve}) achieved in our high-contrast imaging of AB~Aur.}
\label{tab:contrast}
\tablehead{
    \colhead{Separation} & \colhead{Spatially Resolved} & \colhead{Unresolved} \\
    \colhead{($''$)} & \colhead{Contrast ($\Delta$mag)} & \colhead{Contrast ($\Delta$mag)}
    }
\startdata
0.26 & -5.22             & -6.48               \\
0.32 & -5.93             & -7.46               \\
0.39 & -7.54             & -8.21               \\
0.45 & -8.07             & -8.75               \\
0.52 & -8.71             & -9.46               \\
0.58 & -9.58             & -9.84               \\
0.65 & -9.83             & -10.13              \\
0.71 & -10.27            & -10.41              \\
0.78 & -10.53            & -10.69              \\
0.84 & -10.78            & -10.95              \\
0.91 & -10.99            & -11.2               \\
0.97 & -11.17            & -11.39              \\
1.04 & -11.31            & -11.53              \\
1.1  & -11.52            & -11.67              \\
1.17 & -11.57            & -11.76              \\
1.23 & -11.69            & -11.82              \\
1.3  & -11.74            & -11.85              \\
1.36 & -11.75            & -11.87              \\
1.43 & -11.82            & -11.9               \\
1.49 & -11.83            & -11.91              \\
1.56 & -11.8             & -11.92              \\
1.62 & -11.82            & -11.94              \\
1.69 & -11.83            & -11.96              \\
1.75 & -11.89            & -12                 \\
1.82 & -11.92            & -12.02              \\
1.88 & -11.94            & -12.05              \\
1.95 & -11.99            & -12.08              \\
2.01 & -12               & -12.09              \\
2.08 & -12.01            & -12.08              \\
2.14 & -12.01            & -12.07              \\
2.21 & -11.98            & -12.05              \\
2.27 & -11.99            & -12.07              \\
2.34 & -11.99            & -12.08              \\
2.4  & -12               & -12.09              \\
2.47 & -11.96            & -12.05              \\
\enddata
\end{deluxetable}

\end{document}